\documentstyle[aps,prc,twocolumn,epsf,epsfig]{revtex}
\newcommand \bea{\begin{eqnarray}}
\newcommand \eea{\end{eqnarray}}
\newcommand \beq{\begin{eqnarray}}
\newcommand \eeq{\end{eqnarray}}

\newcommand{\ve}[1]{\mbox{\boldmath $#1$}}
\begin{document}
\twocolumn[\hsize\textwidth\columnwidth\hsize
\csname@twocolumnfalse%
\endcsname
\draft
\title{Multipair contributions to the spin response of
nuclear matter} \author{{E.\ Olsson$^{1,2,3}$} and {C.\ J.\
Pethick$^{2,3}$}} \address{
${^1}$ Department of Astronomy and Space Physics, Uppsala University,
 Box 515,
SE-75120 Uppsala, Sweden\\
$^{2}$ NORDITA, Blegdamsvej 17, DK-2100 Copenhagen \O, Denmark\\
$^{3}$ Institute for Nuclear Theory, University of Washington, Box
 351550,
Seattle WA 98195-1550\\}
\maketitle

\begin{abstract}
We analyse the effect of non-central forces on the magnetic
susceptibility of degenerate Fermi systems.  These include
the presence of contributions from transitions
to states containing more than one
quasiparticle-quasihole pair, which cannot be
calculated within the framework of Landau Fermi-liquid
theory, and renormalization of the quasiparticle magnetic moment,
as well as explicit non-central contributions to the quasiparticle
interaction. Consequently, the relationship between the Landau
parameters and the magnetic susceptibility for Fermi systems with
non-central forces is considerably more complicated than for systems
with central forces. We use sum-rule arguments to place a lower bound
on the contribution
to the static susceptibility coming from transitions to multipair
states.


\end{abstract}

\vskip1pc]


\section{Introduction}
The spin and spin-isospin responses of nuclear matter with various
 ratios
of protons to neutrons are important ingredients in many calculations
of importance for astrophysical applications.  Among these we may
mention neutral-current processes such as scattering of neutrinos, and
emission  of neutrino-antineutrino pairs, and charged current ones such
as the emission of neutrinos and antineutrinos by variants of the Urca
process. For a general review of neutrino processes in dense matter, we
refer to Ref. \cite{prakash}.

In calculating rates of neutrino emission, absorption and scattering
processes in dense matter, one of the quantities of interest is the
effective interaction between nucleons, because this is one of
the quantities which determines how matrix elements of the weak current
are modified by the medium (see, e.g., Refs.\ \cite{sawyer,iwamoto}).
One approach to determining these is to calculate effective interactions
directly as, for example, in Ref.\cite{schwenk}. Another approach, which
has been very successful for liquid $^{3}$He is to make use of results
for the long-wavelength thermodynamic properties of the system. To be
specific, in a system with central interactions, some information about
the effective interaction may be obtained rather directly from the spin
susceptibility.  For the purpose of illustration, let us consider such a
system with a single species, having spin $1/2$. The static,
long-wavelength spin susceptibility is given by \beq \chi=\frac{\mu_0^2
N(0)}{1+G_0}, \label{susc0} \eeq
where $\mu_0$ is the magnetic moment of a particle in free space,
$G_0$ is the isotropic part of the Landau quasiparticle
interaction in the spin-triplet channel, and $N(0)=m^* p_{\rm F}/\pi^2
\hbar^3$ is the density of states per unit volume at the Fermi surface,
 $p_{\rm F}$
being the Fermi momentum and $m^*$ the effective mass of a
quasiparticle. Thus, knowing the magnetic susceptibility
and the quasiparticle effective mass, one can deduce the Landau
parameter $G_0$.

Landau's theory of normal Fermi liquids provides a framework for
 describing
the long-wavelength response of the system \cite{landau}. In the usual
 account
of the theory, a
key role is played by conservation laws.  For example,  for interactions
 which
conserve the number of particles, addition of a single quasiparticle to
 the
system involves the addition of a single particle.  The total number  of
particles in the system is therefore given by the total number of
quasiparticles. For interactions which conserve the total spin, the
 total spin may likewise be calculated in terms of the quasiparticle
distribution,  since the spin carried by a quasiparticle is the same as
that carried by a  particle.

The consequences of conservation laws for matrix elements of the density
operator have been described by Pines and Nozi\`eres \cite{pines}, and
 Leggett
has exploited conservation laws to provide a general description of the
long-wavelength response in terms of contributions due to excitation of
 single
quasiparticle-quasihole pairs, and those due to multipair excitations
\cite{leggett}. A derivation of the results from Landau theory is given
 in Ref.\
\cite{baym}.

The assumption that interparticle interactions conserve the total spin
 is an
excellent approximation for atomic interactions, since the spin-orbit
 and
magnetic dipole--dipole interactions between electronic spins, which do
 not
conserve total spin, are generally small.
The situation is quite different for nuclear forces, because of the
 importance
of the tensor interaction due to exchange of pions and, to a lesser
 extent, of
rho mesons, and the significant spin-orbit
contributions to inter-nucleon interactions. For recent analyses of the
nucleon-nucleon interaction, see Refs.\cite{nnint}.

The purpose of this paper is to investigate the consequences of the
lack of conservation of total spin for the spin
response of nuclear matter at long wavelengths.  One of them is
that the spin
carried by a quasiparticle is different from that carried by a bare
 nucleon.
Consequently the coupling of the quasiparticle spin to an external field
 is
altered. Such effects are well known for nuclear magnetic moments and
for matrix elements of the axial current, and  they are
 reviewed in, e.\ g.\ , Ref. \cite{arima}.  They are also discussed in
the  framework of Landau Fermi-liquid theory in Ref.\ \cite{migdal}.
A related effect is that the quasiparticle magnetic moment is a
tensor, not a scalar. A second well-known effect is that the
interaction between quasiparticles has non-central contributions, and
the most familiar of these is that due to the tensor force. This has
been considered in Ref.\cite{haensel} and more recent works.
A
third consequence of non-central forces is that at long wavelengths
there are contributions to the response from states with more than one
quasiparticle-quasihole pair. This latter effect appears not to have
been widely appreciated in work on bulk nuclear matter. In this paper we
examine the general structure of the spin-density response function and
deduce the qualitative behavior of matrix
elements at long wavelengths.  We also indicate how the standard
Landau-theory expression (\ref{susc0}) must be modified when non-central
forces are present. Finally, we derive an upper bound on the magnitude
of the multipair contribution to the long-wavelength susceptibility from
sum-rule arguments.

This paper is organized as follows.  In Section \ref{Sec:response} we
 give 
a brief
introduction to response functions and describe quantities relevant for
our later considerations.  Section \ref{Sec:conservation} gives a
 discussion of the
long-wavelength behavior of matrix elements, and how this depends on
whether or not the corresponding operator obeys a local conservation
law. The consequences of non-central forces for Landau's theory of a
normal Fermi liquid is examined in Section \ref{Sec:landau}.  
In Section \ref{Sec:multipair}, we derive
bounds on the contribution from multipair states to the
static, long-wavelength spin susceptibility and use microscopic
calculations to estimate how large these are.  Concluding remarks are
given in Section \ref{Sec:conclusions}.

\section{Response functions and sum rules}
\label{Sec:response}

If a system initially in its ground state is subjected to a perturbation
\beq
H'=  {\cal O}^{\dagger}_{\bf q} C_{\bf q}e^{-i\omega t} +{\rm Hermitean\
 conjugate},
\label{pert}
\eeq
the Fourier transform of the linear response of the expectation value of
 the
operator ${\cal O}_{\bf q}$ at frequency $\omega$ is given by
\cite{pines2} \beq
<{\cal O}_{\bf q}>_{\omega} = \chi(q, \omega) C_{\bf q},
\eeq
where the linear response function is defined by
\beq
\chi(q, \omega)=\sum_j  |{{\cal O}_{\bf q}^{\dagger}}|^2_{j0}
\frac{2\omega_{j0}}{{\omega}_{j0}^2-(\omega+i\eta)^2}.
\label{response}
\eeq
Here $j$ denotes an excited state and 0 the ground state, and
 $\omega_{j0}=E_j -
E_0$, where $E_j$ is the energy of the excited state, and $E_0$ that of
 the
ground state. The wave vector of the perturbation is denoted by ${\bf
 q}/\hbar$.
 From Eq.\ (\ref{response}) it follows that the static
 response
function is given by
\beq
\chi(q, 0)=\sum_j  \frac{2|{\cal O}_{\bf
 q}^{\dagger}|^2_{j0}}{{\omega}_{j0}}.
\label{staticresponse}
\eeq
Another important quantity is the static structure factor
\beq
S({\bf q}) =\frac{1}{N} \sum_j  |{\cal O}_{\bf q}^{\dagger}|^2_{j0}.
\label{staticstructure}
\eeq
Here $N$ is the total number of particles.
A third quantity we shall find useful, because it can be calculated
 directly
from knowledge of the ground-state wave function, is the
 frequency-weighted sum
\beq
W({\bf q}) =\frac{1}{N} \sum_j  |{\cal O}_{\bf
q}^{\dagger}|^2_{j0}{\omega}_{j0} = \frac{1}{N}\langle 0| {\cal
O}_{\bf q}[H, {\cal
O}^{\dagger}_{\bf q}]|0\rangle,
\eeq
where $H$ is the Hamiltonian in the absence of the perturbation given in
Eq.(\ref{pert}).  As we shall demonstrate later, the static structure
factor and frequency-weighted sum at long-wavelengths are useful
diagnostics for contributions from multipair states.

\section{Conservation laws and non-central forces}
\label{Sec:conservation}
We now investigate the effect of non-central components of the
 nuclear force on the linear response of the system.
One of the reasons for the spectacular success of Landau Fermi-liquid
 theory in
providing a framework for the quantitative description of
 long-wavelength
low-frequency properties of liquid $^3$He at low temperatures is that
 the
physical observables of greatest interest experimentally correspond to
quantities such as the particle density, the spin density, and the
 particle
current density, which satisfy local conservation laws.  To make this
 point
explicit, let us imagine that the operator ${\cal O}(\bf r)$ satisfies
 the
conservation law
\beq
\frac{\partial {\cal O}({\bf r})}{\partial t} +\ve{\nabla}\cdot{\bf
 j}({\bf r})
=0,
\eeq
where ${\bf j}({\bf r})$ is the associated current density.  On taking
the matrix element of the Hermitian conjugate of this equation between
 an excited state $j$
and the ground state and Fourier transforming in space, one finds
\beq
\omega_{j0}({\cal O}^{\dagger}_{\bf q})_{j0}=
{\bf q}\cdot({\bf j}^{\dagger}_{\bf q})_{j0}.
\eeq
This shows that the matrix element of the operator satisfies the
 condition
\beq
({\cal O}^{\dagger}_{\bf q})_{j0}=
\frac{{\bf q}\cdot({\bf j}^{\dagger}_{\bf q})_{j0}}{\omega_{j0}}.
\label{matrixelement}
\eeq
Thus for a state for which $\omega_{j0}$ tends to a nonzero value as
$q$ tends to zero,
the matrix element of the operator tends
to zero in this limit, provided only that the matrix element of
 the
current remains finite.

The expression (\ref{matrixelement}) provides the basis for an
 investigation of
contributions to physical quantities coming from single-pair states and
multipair ones.  As an example, let us consider the case of density
fluctuations, that is ${\cal O}_{\bf q}$ is equal to the Fourier
 transform of
the particle density operator $\rho_{\bf q}$.  For a single-pair state,
 with a
quasihole in the state with momentum ${\bf p + q}$ and a quasiparticle
 of
 momentum
${\bf p}$, the matrix element of the current operator at long
wavelengths is proportional to ${\bf p}/m$, where $m$ is the mass of the
particle.  Since the excitation energy of a single pair is
 $\omega_{j0}\sim
\epsilon_{\bf p+q}- \epsilon_{\bf p}= {\bf v}_{\bf p}.{\bf q}$, where $
 {\bf
v}_{\bf p}=  \ve{\nabla}_{\bf p}\epsilon_{\bf p}$, the matrix element of
 the density operator tends to a constant.

Let us now examine the contributions of
single-pair states to the static density-density response function,
the static structure factor, and the frequency-weighted sum. In a
degenerate Fermi system at zero temperature, the number of  states
with a single quasiparticle-quasihole pair having total momentum $\bf q$
is of order $nq/p_{\rm F}$ per unit volume for $q \ll p_{\rm F}$.   The
 typical energy of a single pair
excitation is of order $v_{\rm F}q$, where
$v_{\rm F}$ is the Fermi velocity, and consequently the contribution
to the static structure factor varies as $q$, that to the static
response function as $q^0$, and the contribution to the 
frequency-weighted sum as $q^2$.  These behaviors are the same as for a
free Fermi gas. Similar results apply for the corresponding response
function for the spin density, irrespective of the nature of the
interaction.  We shall not discuss possible collective modes here, but
the matrix elements of the density operator to such states, which have
energies $sq$, where $s$ is the velocity of the collective mode, may
be shown to vary as $q^{1/2}$.  Thus the contributions to
$\chi(q, 0)$, $S(q)$ and $W(q)$ have the same dependence on $q$ as those
from the single-pair states.

Now let us turn to the multipair contributions.  As Pines and Nozi\`eres
demonstrated \cite{pines2}, from Eq. (\ref{matrixelement}) it follows
 that the matrix
element of the density operator vanishes at least as rapidly as $q$ for
$q \rightarrow 0$.  If the system is translationally invariant, the
ground state of the system is an eigenstate of the total current,
and from this it follows that matrix elements of the current between
states which are orthogonal in the limit $q \rightarrow 0$ must vanish
in that limit.  If $({\bf j}_{\bf q})_{j0}$ can be expanded in a Taylor
series, this implies that $(\rho_{\bf q})_{j0}$ varies as $q^2$ for
small $q$.  The density of multipair states does not depend crucially on
$q$ for small $q$, and therefore it follows that the contributions of
multipair excitations to $\chi(q, 0)$, $S(q)$ and $W(q)$ vary as $q^4$
for translationally invariant systems, and as $q^2$ for systems without
translational invariance.

Now let us turn to the spin response.  If the interaction conserves
total spin, the contributions to the moments of the spin-spin response
function have the same behavior as for the density. The multipair
contributions vary as $q^2$, since the spin current is not generally
conserved, except for very special forms of the interaction. For
nucleons, the interaction has non-central components and therefore spin
is not conserved.  Consequently, matrix elements of the spin
density operator do not vanish in the limit $q\rightarrow 0$, and one
arrives at the striking conclusion that the contributions to $\chi(q,
0)$, $S(q)$ and $W(q)$ are all nonzero for $q \rightarrow 0$.  In the
following section we shall examine the physical content of this result
in terms of Landau's theory of normal Fermi liquids.

\section{Landau theory}
\label{Sec:landau}
In this section we discuss the long-wavelength low-frequency response of
 the
system within the framework of Landau Fermi-liquid theory.  The basic
 ideas
have been presented in the paper by Leggett \cite{leggett}, and we shall
extend his treatment to the case of tensor forces.  While Leggett's work
 was
couched in terms of field-theoretical methods, we shall follow more
closely the phenomenological approach of Landau.

Haensel and D\c{a}browski have shown that for a magnetically polarized
 Fermi
liquid, the tensor force causes the Fermi surface to become aspherical
\cite{haensel}. In this paper we shall confine our attention to small
polarizations, so the distortions of the Fermi surface may be
treated as small.

In a magnetically polarized Fermi system the quasiparticle
distribution function is a matrix $(n_{\bf p})_{\alpha \beta}$ in
spin space.  The quasiparticle energy, which is likewise a matrix in
spin space, is defined as the functional derivative of the total
energy with respect to the quasiparticle distribution:
\beq
\delta E[n_{\bf p}]=\sum_{\alpha \beta}(\epsilon_{\bf p})_{\alpha
\beta}(\delta n_{\bf p})_{\beta \alpha}.
\label{epsilon}
\eeq
The quasiparticle interaction is a matrix in two pairs of spin
indices, and it is given as the second functional derivative of the
energy with respect to the distribution function
 \beq
\delta^2 E[n_{\bf p}]=\frac{1}{2}\sum_{\alpha \beta \alpha' \beta'}
(f_{{\bf p},{\bf p'}})_{\alpha \beta, \alpha' \beta'}(\delta n_{\bf p})_
{\beta\alpha} (\delta n_{\bf p'})_{\beta'\alpha'}.
\label{f}
\eeq

The magnetic moment of a quasiparticle is related to the dependence
of the quasiparticle energy on the  magnetic field ${\ve {\mathcal H}}$.
It is convenient to write the quasiparticle energy in the form
\beq
(\epsilon_{\bf p})_{\alpha \beta}=\epsilon_{\bf p}\delta_{\alpha \beta}
 + \sum_{i=1}^3\epsilon^i_{\bf p}{\sigma}^i_{\alpha \beta},
\eeq
where the ${\sigma}^i$ are the Pauli matrices.  When non-central forces
are present, application of a magnetic field will tend
to produce a polarization of the quasiparticle distribution for a
particular momentum in a direction different from that of the applied
field. Thus the magnetic moment of a quasiparticle is a tensor, and it
is given by
\beq \mu_{ij}({\bf p})=- \frac{\partial \epsilon^i_{\bf
p}}{\partial {\mathcal H}_j}.
\eeq
If only central forces act, and one neglects excitation of degrees of
freedom other than the nucleons, the total spin is conserved, and the
magnetic moment is proportional to the component of the spin in the
direction $i$. It then follows from
conservation of the component of the total spin in the
direction of the magnetic field, that the magnetic moment of a
quasiparticle in a single-component Fermi system is equal to the moment
of a bare particle.

The form of the magnetic moment tensor when non-central forces are
present may  be
written down immediately from the requirement that it be
 invariant under
simultaneous rotations of the directions of the magnetic field and of
 the
momentum of the quasiparticle.  This dictates that the expression for
 the
magnetic moment is
\beq
\mu_{ij}({\bf p})=\mu \delta_{ij} +\frac{3\mu_T}{2} \left(\frac{p_i p_j}{p^2}
 -\frac{\delta_{ij}}{3}\right).
\label{Eq:magneticmoment}
\eeq
Here
$\mu$, which is not generally equal to the bare moment $\mu_0$, is
the magnetic moment averaged over directions of the quasiparticle
momentum and
$\mu_T$ is a coefficient characterizing the strength of the
off-diagonal part of the magnetic moment tensor. For
the specific case of a one-pion-exchange potential this form was derived
long ago by Miyazawa \cite{miyazawa}. The general form of the
quasiparticle interaction may be determined by symmetry considerations.
For purely central interactions, and for an unpolarized system the
interaction must be invariant under rotation of the spins of the two
quasiparticles, and therefore the interaction must be of the standard
exchange form 
\beq 
f_{{\bf p} \ve{\scriptstyle{\sigma}},{\bf
p'}\ve{\scriptstyle{\sigma'}}}=f_{{\bf p}{\bf p'}}+g_{{\bf p}{\bf
p'}}\ve{\sigma}\cdot \ve{\sigma'}, 
\eeq
where we have used a compact
matrix notation as usual. With the matrix indices included, this
equation is 
\bea
\nonumber
(f_{{\bf p }\ve{\scriptstyle{\sigma}},{\bf
 p'}\ve{\scriptstyle{\sigma'}}})_{\alpha \beta,\alpha' \beta'}&=&f_{{\bf
 p}{\bf p'}}\delta _{\alpha \beta} \delta _{\alpha' \beta'}+g_{{\bf p}
 {\bf p'}}({\ve \sigma})_{\alpha \beta}\cdot({\ve \sigma'})_{\alpha'
 \beta'}.\\
\eea
For calculating low-temperature properties, the quasiparticle
 interaction 
is needed only for $p$ and $p'$ equal to the Fermi momentum $p_{\rm F}$,
and  in this case $f$ and $g$ are functions only of the angle
 between ${\bf p}$ and ${\bf p'}$, which we denote by $\theta$.

When non-central forces are present, the interaction is not invariant
under rotations of the spins and the momenta separately, but only under
simultaneous rotations of spins and momenta.  Then  there can be
additional terms of the form ${\ve \sigma}\!\cdot\!{\bf p}\
 {\ve\sigma'}\!\cdot\!{\bf p}$, and other terms with one or both of the
 momenta
$\bf p$ in this expression replaced by ${\bf p}'$. For nuclear  matter,
 the most important contribution to the non-central part of the
 interaction comes from pion exchange, so it is generally assumed that
 the interaction may be taken to be of the form suggested by the
 result one finds for one-pion exchange
\bea
\nonumber
f_{{\bf p} \ve{\scriptstyle\sigma},{\bf p'}  \ve{\scriptstyle\sigma}'}
&=&
f_{{\bf p} {\bf p'}}  +
g_{{\bf p} {\bf p'}}{\ve{\sigma}}\!\cdot\!{\ve{\sigma}'}
+  k_{{\bf p} {\bf p'}}(3{\ve{\sigma}}\!\cdot\!\hat{\bf
u}\ {\ve{\sigma}'}\!\cdot\!\hat{\bf u}-
{\ve{\sigma}}\!\cdot\!{\ve{\sigma}'}),\\
\eea
where  ${\bf u}={\bf p}-{\bf p'}$.
As for $f_{{\bf p} {\bf p'}}$ and $g_{{\bf p} {\bf p'}}$ the function
 $k_{{\bf p} {\bf p'}}$ is also a function only of
 $\theta$ at low temperatures. In earlier work, e.g. Refs.
\cite{haensel,dabrowski}, the non-central contribution to the
quasiparticle interaction has usually been expressed in terms of a
function $h$ which is related to $k$ by the equation $k=hu^2/p_{\rm
F}^2$. As we shall explain elsewhere, the advantage of the
parametrization we use is that the expansion of $k$ in terms of Legendre
polynomials converges much more rapidly than that for $h$ \cite{olsson}.
Therefore,  all of these functions can be
expanded in terms of Legendre polynomials of $\cos\theta$.  For example,
for $k$ we write 
\beq k_{{\bf p} {\bf p'}}=\sum_{l=0}^{\infty}k_l P_l
(\cos \theta). \eeq

Now we quote results for the static magnetic susceptibility.
 Previously,
Haensel and Jerzak \cite{jerzak} performed a similar calculation
including the tensor contribution to the quasiparticle interaction but
neglecting multipair contributions and the renormalization of
the magnetic moment.
The magnetization of the matter is
\beq
M_i = \sum_{p} {\rm Tr} \mu_{ij} \sigma_j \delta n_{\bf p} + M_{\rm M},
\eeq
where $M_{\rm M}$ is the contribution to the magnetization coming from
 multipair excitations.
And the spin susceptibility is defined as
\beq
\chi = \frac{\partial M_i}{\partial {\mathcal H}_i}\bigg|_{{\mathcal
 H}=0}.
\eeq
For simplicity, we here give the result obtained if
one includes only Landau  parameters
with $l<2$ for the central part of the interaction, and takes only the
 $l=0$
term in the tensor part of the interaction $k$.
Using  Eq. (\ref{Eq:magneticmoment})
 for the magnetic moment and
including the tensor interaction up to second order,
we arrive at the following expression for the static spin
susceptibility:
 \bea
\nonumber
\chi&=&
\frac{{\mu}^2 N(0)}{1+G_0-{K_0}^2/8}-\frac{\mu \mu_T N(0)
 K_0}{1+G_0}+\frac{1}{2}\ N(0){\mu_T}^2 +\chi_{\rm M},\\
\label{eq:chi}
\eea
The corresponding result if renormalization of the magnetic moment and
multipair contributions are neglected is
$\chi=\mu_0^2N(0)/(1+G_0-{K_0}^2/8)$. More details about the above
calculation and the choice of  parametrization will be given in a later
paper, \cite{olsson}. Next we
shall obtain a lower bound on the magnitude of the last term in
Eq. (\ref{eq:chi}), the  contribution from multipair excitations.

\section{A bound on multipair contributions to the spin susceptibility}
\label{Sec:multipair}
In this section we shall consider the part of the static susceptibility
 that
cannot be calculated within Landau theory, the contributions from
 multipair
excitations. Our approach will be to employ sum-rule arguments to place
a lower bound on the quantity.
Since the excitation energies of the system are positive, it
follows for any choice of the energy $\Omega$ that
 \beq
 \sum_j |<j|{\cal{O}}_{\bf q}|0>|^2\frac{(\Omega -
 \omega_{j0})^2}{\omega_{j0}} \ge 0.
\eeq
If one takes
 $\Omega$ to be the mean excitation energy ${\bar \omega}
 =W(q)/S(q)$, this yields the inequality
\beq
\chi(q, 0)\ge  \frac{2 n S(q)}{\bar \omega},
\eeq
which is valid for all $q$.  Let us now apply this result in the limit
$q\rightarrow 0$.  As we argued in Sec.\ \ref{Sec:conservation},
single-pair excitations and collective modes do not contribute to $\chi$
in this limit, and therefore it follows that
\beq
\chi_{\rm M}(q, 0)\ge  \frac{2 n S(q)}{\bar \omega} \qquad (q\rightarrow
0),
 \label{chiM}
\eeq
where $\chi_{\rm M}(q, 0)$ is the multipair part of the static response
 defined
in Eq. (\ref{staticresponse}).

We now apply this result to pure neutrons, using values of the static
structure factor and the mean excitation energy
in the limit $q\rightarrow 0$  taken from Ref.\cite{akmal}.
In that reference the authors calculated the spin-spin response
function, while in this paper we have worked in terms of the
response of the magnetization density.  The response functions we need
are therefore $\mu_0^2$ times those of Ref.\ \cite{akmal}.

 At the saturation density of nuclear matter, $n=0.16\
{\rm fm}^{-3}$, these are $S_{\rm M}(q\!\!=\!\!0)\approx 0.19 \mu_0^2$ and $\bar
\omega(q\!\!=\!\!0) \approx 63 \ {\rm  MeV}$. For the total
 susceptibility
we take the value  $\chi\approx 0.38 \chi_{\rm F}$   from the recent calculations of Fantoni {\it et
al.}  \cite{fantoni} using the auxiliary field diffusion Monte Carlo
method.
Here $\chi_{\rm F}\!=\!3 n \mu_0^2/ 2\epsilon_{\rm F}$ is the
susceptibility of a free Fermi gas.  The quantity $\epsilon_{\rm F}$ is
the Fermi energy of the non-interacting gas, and at a density $n=0.16\
{\rm fm}^{-3}$ this is approximately $59\ {\rm MeV}$ for neutrons.
The value for the susceptibility
obtained in Ref.\ \cite{fantoni} is close to the results of earlier
calculations which included fewer correlations,
(See e.g. Ref.\ \cite{bj}). Thus from Eq.\ (\ref{chiM}) we find
${\chi_{\rm M}(q, 0)}/{{\chi}}\ge 0.63$.
If this result were to be taken at face value, it would indicate
that multipair states contribute more than $\sim$
60\% of the total static response function. This would appear to be
unrealistically high.  The calculations of Ref.\cite{akmal}
were designed primarily to investigate the spin response at
finite wavelengths, at which phenomena such as pion
condensation would be expected to occur.  Estimating
long-wavelength response is a difficult problem because it
requires a careful investigation of long-range correlations.
 The difficulty may be illustrated by considering experience
with calculations of the static structure factor for density response
for liquid $^4$He.  On the basis of general arguments of the sort given
above, the static structure factor should vanish in the limit
$q\rightarrow 0$.  However, early calculations gave structure factors
which were nonzero in this limit.  Our conclusion is that sum rule
arguments of the sort given above can in principle provide valuable
information about multipair states, but that more detailed calculations
of structure factors are needed before they can provide useful
quantitative estimates.

\section{Conclusions}
\label{Sec:conclusions}

One of the main conclusions of this paper is that, when
non-central forces are present, there are contributions to the static
magnetic susceptibility at long wavelengths that cannot be calculated
within the framework of Landau Fermi-liquid theory.  In addition,
non-central forces renormalize the magnetic moment of a quasiparticle,
and introduce additional terms in the expression for the effective
interaction between quasiparticles.  Consequently, it is much less
straightforward to obtain information about Landau parameters from the
magnetic susceptibility than it is when there are only central forces.
This may be seen by comparing Eqs.\ (\ref{susc0}) and (\ref{eq:chi}).

Among quantities which need to be understood better in order to
relate quasiparticle interactions to the magnetic susceptibility are
the strength of transitions to multipair states and the renormalization
of the magnetic moment of a quasiparticle. In order to place more
reliable bounds on the strength of transitions to multipair excitations
it would be valuable to have improved estimates of the static spin
structure factor at long wavelengths.  The renormalization of the
magnetic moment has been studied recently by Cowell and Pandharipande
using the correlated basis approach developed from the variational
methods used for quantum liquids \cite{cowell}. They find that the
magnitudes of magnetic moments are reduced by approximately 10\% for nuclear matter
with proton fractions ranging from 0.2 - 0.5 and for densities between
0.08 fm$^{-3}$ and 0.24 fm$^{-3}$.

In this paper we have focussed mainly on a single-component system of
fermions with spin-$1/2$.  Our arguments may be extended
straightforwardly to multi-component systems, such as nuclear matter with
an arbitrary ratio of neutrons to protons. In addition, similar
arguments may be applied for the spin-isospin response.

\section{Acknowledgements}   We are grateful to Arya Akmal for giving us the
data for the static structure factor and the energy-weighted response.
We also thank Bengt Friman, Wick Haxton, Ben Mottelson, Vijay Pandharipande 
and the participants in the program on ``Neutron stars'' at the Institute for
Nuclear Theory for valuable discussions.  One of us (EO) acknowledges
financial support in part from a European Commission Marie Curie Training 
Site Fellowship under Contract No. HPMT-2000-00100. 
This work has been conducted within the framework of the school on
Advanced Instrumentation and Measurements (AIM) at Uppsala University
supported financially by the Foundation for Strategic Research (SSF).

\end{document}